\documentclass[prd,nofootinbib,aps,10pt,superscriptaddress,article,twocolumn]{revtex4-1}

\usepackage{amsmath,amsfonts,amssymb,amsthm}
\usepackage{mathtools}
\usepackage{setspace}
\usepackage[normalem]{ulem}
\usepackage{stmaryrd}
\usepackage{expdlist}

\usepackage{graphicx,xcolor}

\usepackage{verbatim}
\usepackage{alltt}
\usepackage{moreverb}

\usepackage{xr}

\usepackage{enumitem}
\usepackage[symbol*,hang,perpage,para]{footmisc}

\newcommand{\kpc}{\text{ kpc}}
\newcommand{\e}{\text{ e}}
\newcommand{\s}{\text{s}}
\newcommand{\m}{\text{ m}}
\newcommand{\cm}{\text{cm}}
\newcommand{\km}{\text{ km}}

\newcommand{\Si}{\text{Si}}
\newcommand{\Ci}{\text{Ci}}
\newcommand{\Bj}{\text{j}}
\newcommand{\By}{\text{y}}
\newcommand{\Y}{\text{Y}}
\newcommand{\LP}{\text{P}}

\begin{document}

\title{Dark-matter-like solutions to Einstein's ``unified" field equations}

\author{J.~R. van Meter}
\email{James.VanMeter@Colorado.edu}
\affiliation{Dept. of Mathematics, University of Colorado, Boulder, CO, 80309}

\begin{abstract}
Einstein originally proposed a nonsymmetric tensor field, with its symmetric
part associated with the spacetime metric and its antisymmetric part associated 
with the electromagnetic field, as an approach to a unified field theory.  Here 
we interpret it more modestly as an alternative to Einstein-Maxwell theory, 
approximating the coupling between the electromagnetic field and spacetime 
curvature in the macroscopic classical regime.  Previously it was shown 
that the Lorentz force can be derived from this theory, albeit with deviation 
on the scale of a universal length constant $\ell$.  Here we assume that $\ell$ is of 
galactic scale and show that the modified coupling of the electromagnetic field 
with charged particles allows a non-Maxwellian equilibrium of 
non-neutral plasma.  The resulting electromagnetic field is ``dark" 
in the sense that its modified Lorentz force on the plasma vanishes, 
yet through its modified coupling 
to the gravitational field it engenders a nonvanishing, effective mass density.
We obtain
a solution for which this mass density asymptotes approximately to that of the pseudo-isothermal model of dark matter.  The resulting gravitational field produces radial 
acceleration,
in the context of a post-Minkowskian approximation, 
which is negligible at small radius but yields a flat rotation curve at large radius. 
We further exhibit a family of such 
solutions which,
like the pseudo-isothermal model, has a free parameter 
to set the mass scale (in this case related to the charge density) 
and a free parameter to set the length scale (in this case an
integer multiple of $\ell$).  
Moreover, these solutions are members of a larger family with more general
angular and radial dependence.  They thus show promise as approximations of generalized
pseudo-isothermal models, which in turn are known to fit a wide range of 
mass density profiles for galaxies and clusters.  
\end{abstract}

\maketitle

\section{Introduction}

The nature of dark matter has resisted scientific consensus since its
initial observation nearly a century ago \cite{1933AcHPh...6..110Z,1937ApJ....86..217Z,Koushiappas:2017chw}.
To date, no strong evidence has been found for 
weakly interacting massive particles,
nor any other hypothesized constituent of dark matter halos \cite{Yang:2016odq,Akerib:2016vxi,Koushiappas:2017chw,PhysRevLett.119.181301,PhysRevLett.119.181302}.
Meanwhile, alternatives to dark matter involving modified theories of gravity 
have failed to gain widespread acceptance, partly due to difficulties 
in fitting all of the empirical data (e.g.~\cite{1992ApJ...397...38G,Sanders:2002ue}).

In view of this long-standing lack of a satisfying theory for dark matter, 
consideration of less conventional explanations may be warranted.
As an alternative to new matter on the one hand, and modified gravity
on the other, we offer a third alternative, that dark matter is actually
conventional matter behaving unconventionally on galactic scales,
within the context of a modified theory of gravity (and electrodynamics).
This explanation is also distinguished from others in that it
is not a phenomenological contrivance, but 
rather emerges naturally from a theory proposed by Einstein,
also nearly a century ago.

Einstein first proposed his theory of a nonsymmetric tensor field in 1925,
intending it to unify the gravitational and electromagnetic fields 
\cite{Einstein:1925:EFG}.
He returned to it in 1945 \cite{Einstein:1945:GRT} and seemed to favor it until his passing ten years
later.  Indeed, it is the subject of his final scientific publication
\cite[Appendix II]{Einstein:1955:MR} (see also \cite{doi:10.1119/1.11809}).

This theory received serious attention during Einstein's lifetime, 
including an independent derivation by Schr\"odinger \cite{schrödinger1950space}.  
Subsequently,
however, Einstein's theory came to be neglected by the physics community at large, for 
reasons we will touch on below.  Notable exceptions include the work of 
Johnson, who beginning in 1971 explored the consistency of Einstein's 
nonsymmetric field theory with conventional electrodynamics and gravitation 
\cite{PhysRevD.4.295,PhysRevD.4.318,PhysRevD.4.3555,PhysRevD.5.282,PhysRevD.5.1916,PhysRevD.5.1916,PhysRevD.7.2825,PhysRevD.7.2838,PhysRevD.8.1645,Johnson1972,PhysRevD.15.377,PhysRevD.24.327,PhysRevD.31.1236,PhysRevD.31.1252,PhysRevD.34.1025}, 
and the work of Moffat and others 
who beginning in 1979 considered Einstein's nonsymmetric field without the 
electromagnetic interpretation \cite{PhysRevD.19.3554}.  We will build upon the former.

Some immediate concerns may come to mind.  First,
Einstein's simplistic geometric approach to unification is widely
regarded as a failure, and misguided from the start, due to its
apparent disregard of all but the classical gravitational and electromagnetic
fields (at best).  In the present
work we do not consider Einstein's nonsymmetric field theory as a unified theory.
Rather, we only apply it to macroscopic scales and in regimes expected
to be well approximated by classical physics.

It is after all an empirical fact that the electromagnetic field couples to the 
gravitational field on scales well below that where unification is expected,
as evidenced by gravitational redshift and gravitational lensing.
Einstein-Maxwell theory, in which Einstein's tensor is set proportional
to the Maxwell stress-energy tensor, adequately explains these effects,
and moreover is generally believed applicable to astrophysical phenomena 
involving strong spacetime curvature and in which the electromagnetic field
plays a significant role (e.g.~\cite{Giacomazzo:2012iv}).  
Coupling as it does the electromagnetic field with the gravitational field
in a manifestly, generally covariant manner, Einstein-Maxwell theory seems
well suited for these tasks.  It is not, however, unique in this regard.
Einstein's nonsymmetric field theory is also generally covariant and, as it happens,
can also be put in a form approximating Einstein's tensor coupled with
an electromagnetic source tensor.  
We argue that Einstein's nonsymmetric field theory be considered viable
wherever Einstein-Maxwell theory currently is.

But this suggestion may raise another concern.
In 1953 it was found by Callaway that, if charged particles are modeled
by a particular form of singularity analogous to that of the Reissner-Nordstr\"om
solution in Einstein-Maxwell theory, then to leading post-Newtonian orders
such a particle in Einstein's nonsymmetric field theory fails to respond
to an electromagnetic field in any way consistent with the Lorentz force
equation \cite{PhysRev.92.1567}.  By contrast, the contracted Bianchi identity applied to the
Maxwell stress-energy tensor, as mandated by Einstein-Maxwell theory,
implies to leading post-Minkowskian order the full Lorentz-Dirac equation
for a Reissner-Nordstr\"om particle
(i.e. the Lorentz force equation with radiative corrections) \cite[\S20.6]{misner1973gravitation}, \cite{PhysRev.57.797,10.2307/2371617,10.2307/20489094}.

This objection was answered straightforwardly by Johnson 
who showed that Callaway did not consider a general enough solution
with which to model a charged particle \cite{PhysRevD.4.318}.
Callaway identified one integration constant of Einstein's
nonsymmetric field equations with mass, and another with charge. 
However, Einstein's third order field equations allow for an 
additional integration constant in a homogeneous monopole solution, 
which Callaway neglected.  Johnson showed that if this extra constant is 
also assumed nonzero, and related to electric charge, 
then the Lorentz-Dirac 
equation can be recovered from 
Einstein's nonsymmetric field theory, to good approximation.

But Johnson's particular identification of integration constants with charge 
comes at the price of a length parameter, necessarily finite, at which scale Coulomb's law breaks 
down.  In the present work we assume this parameter is of extrasolar scale 
($\gtrsim 10^{11}\m$) and, following a suggestion also made by Johnson
that this may lead to an explanation for dark matter \cite{johnson}, 
consider galactic scale charge distributions.   
We then derive a family of post-Minkowskian solutions 
which, we show, exhibit dark-matter-like properties.  The solutions are 
axially symmetric, but at large radius yield flat rotation curves that 
asymptote to spherical symmetry.  

This paper is organized as follows.  In Sec.~II we discuss Einstein's 
original field equations as well as their leading order post-Minkowskian 
expansion.  In Sec.~III.A, we derive the equation 
that a charge distribution needs to satisfy in order to be in electrostatic 
equilibrium, in the context of this theory, and obtain a general solution.  
In Sec.~III.B, we solve for the gravitational field at large radius, 
obtaining particular solutions that yield flat rotation curves.  In 
Sec.~III.C, we show that this family of solutions is valid at all radii, and 
that it comes with a free parameter that determines a ``core radius" analogous 
to the length scaling parameter common to phenomenological dark matter models.  
In Sec.~III.D we show that its gravitational and electric fields are 
negligible near the galactic center and thus consistent with observations of 
an uncharged central black hole.  Finally conclusions are given in Sec.~IV.

\section{Background}

Extremizing the Einstein-Hilbert action,
\begin{equation}
\delta\int d^4x\sqrt{-g}g^{\mu\nu}R_{\mu\nu}=0,
\end{equation}
by standard methods but allowing both $g^{\mu\nu}$ and
$R_{\mu\nu}$ to be asymmetric results in the following equations\cite{Einstein:1955:MR,schrödinger1950space}:
\begin{eqnarray}
R_{(\mu\nu)} &=& 0,\\
R_{[\mu\nu,\rho]} &=& 0,\\
\Gamma^\nu_{[\mu\nu]} &=& 0,
\end{eqnarray}
where $\Gamma^\lambda_{\mu\nu}$ is defined by
\begin{equation}
g_{\sigma\nu}\Gamma^\sigma_{\mu\rho}+g_{\mu\sigma}\Gamma^\sigma_{\rho\nu} = g_{\mu\nu,\rho}
\end{equation}
and $R_{\mu\nu}$ via
\begin{equation}
R_{\mu\nu} = \Gamma^\rho_{\mu\nu,\rho}-\Gamma^\rho_{\mu\rho,\nu}-\Gamma^\rho_{\mu\sigma}\Gamma^\sigma_{\rho\nu}+\Gamma^\rho_{\mu\nu}\Gamma^\sigma_{\rho\sigma}.
\end{equation}
Note that both $R_{\mu\nu}$ and $\Gamma^\lambda_{[\mu\nu]}$ transform
as tensors even though $\Gamma^\lambda_{\mu\nu}$ does not.
Clearly these equations reduce to those of conventional general relativity
when the antisymmetric components vanish.

For the purpose of post-Minkowskian approximation it is convenient to 
define 
\begin{equation}
\gamma^{\mu\nu}\equiv\sqrt{-g}g^{\mu\nu}-\eta^{\mu\nu}.
\end{equation}
Then using the identity \cite{Einstein:1950:BIG},
\begin{equation}
\gamma^{[\mu\nu]}_{\ \ \ \ ,\nu}=\sqrt{-g}g^{(\mu\nu)}\Gamma^\rho_{[\nu\rho]},
\end{equation}
choosing the harmonic gauge condition 
\begin{equation}
\gamma^{(\mu\nu)}_{\ \ \ \ ,\nu}=0, 
\end{equation}
and further defining 
\begin{equation}
\gamma^*_{[\mu\nu]}=\frac{1}{2}\epsilon_{\mu\nu\rho\sigma}\gamma^{[\rho\sigma]},
\end{equation}
we obtain \cite{PhysRevD.4.295}
\begin{eqnarray}
\Box j_\mu &=& s_\mu,\\
j_\mu^{,\mu} &=& 0,\\
\gamma^{*,\nu}_{[\mu\nu]} &=& j_\mu,\\
\gamma^*_{[\mu\nu,\rho]} &=& 0,\\
\Box\gamma_{(\mu\nu)} &=& t_{\mu\nu},
\end{eqnarray}
where $s_\mu=-\frac{1}{3}\eta_{\mu\rho}\epsilon^{\rho\sigma\kappa\lambda}G^N_{[\kappa\lambda,\sigma]}$, $j_\mu$ can be considered a dummy variable introduced to
write third order equations as a second order system, $t_{\mu\nu}=-2G^N_{(\mu\nu)}$, and 
$G^N_{\mu\nu}$ is equal to the nonlinear terms of $R_{\mu\nu}-\frac{1}{2}\eta_{\mu\nu}\eta^{\rho\sigma}R_{\rho\sigma}$.  To leading post-Minkowskian order \cite{PhysRevD.4.318},
\begin{eqnarray}
t_{\mu\nu}&=&\frac{1}{2}\gamma_{[\rho\sigma],\mu}\gamma^{[\rho\sigma]}_{,\nu}+\gamma_{[\mu\rho]}^{,\sigma}\gamma_{[\nu\sigma]}^{,\rho}-\gamma_{[\mu\rho],\sigma}\gamma_{[\nu}^{\rho],\sigma}\nonumber\\
&&-\frac{1}{4}\eta_{\mu\nu}\gamma_{[\rho\sigma],\kappa}\gamma^{[\rho\sigma],\kappa}-\frac{1}{2}\eta_{\mu\nu}\gamma_{[\rho\sigma],\kappa}\gamma^{[\rho\kappa],\sigma}\\
&&+\gamma^{[\rho\sigma]}\gamma_{[\mu\sigma],\nu\rho}+\gamma^{[\rho\sigma]}\gamma_{[\nu\sigma],\mu\rho}-\frac{1}{2}\eta_{\mu\nu}\gamma^{[\rho\sigma]}\Box\gamma_{[\rho\sigma]}.\nonumber
\end{eqnarray}

In \cite{PhysRevD.4.318} it was shown that we can write $j_\mu=J_\mu+A_\mu$, where
$J_\mu$ can be identified with the conventional Maxwell current 
provided we introduce electromagnetic field tensors $F^L_{\mu\nu}$
and $F^D_{\mu\nu}$ (labeled according to their ``local" and ``diffuse" sources,
respectively) defined such that
\begin{equation} 
\label{F2gammastar}
\gamma^*_{[\mu\nu]}=c^{-2}G^{1/2}k^{-1}(F^L_{\mu\nu}+2F^D_{\mu\nu}) 
\end{equation}
and 
\begin{eqnarray}
\label{maxwell}
F^L_{\mu\nu} &=& \partial_\mu A_\nu-\partial_\nu A_\mu,\\
F_{\mu\nu}^{L,\nu} &=& J_\mu,\\
F_{\mu\nu}^{D,\nu} &=& -k^2A_\mu,
\label{maxwelllast}
\end{eqnarray}
where $k=(\sqrt{2}\ell)^{-1}$ and $\ell$ is a universal constant
with units of distance.
(And unless otherwise noted we will use Gaussian units.)
In the Lorenz gauge, then, $A_\mu$ plays the dual role of vector potential
for $F^L_{\mu\nu}$ and source for $F^D_{\mu\nu}$.
With the above identifications and conditions, if $J_\mu$ includes a point
source with mass $m$ and charge $q$ then it was shown that it satisfies the Lorentz-Dirac equation:
\begin{equation}
\label{lorentz}
ma_\mu=qc^{-1}u^\lambda(F^{L\text{ ext}}_{\lambda\mu}+F^{D\text{ ext}}_{\lambda\mu})+\frac{2}{3}q^2c^{-3}\left(\frac{d}{d\tau} a_\mu+a_\rho a^\rho u_\mu\right)
\end{equation}
Clearly if $\ell$ is very large relative to 
the scale of spacetime variation of $F^L_{\mu\nu}$, 
and other length scales of a physical system, 
so that $k^2$ is relatively small, then
Eqs.~(\ref{maxwell}-\ref{lorentz}) reduce to those of conventional electrodynamics.

It should be mentioned that in writing down the above equations we have made particular choices
in how to identify certain physical quantities, e.g. electric charge, 
with quantities in Einstein's nonsymmetric field theory.
We are following, specifically, \cite{PhysRevD.4.318}.
Other choices are explored in 
\cite{PhysRevD.4.295,PhysRevD.4.318,PhysRevD.4.3555,PhysRevD.5.282,PhysRevD.5.1916,PhysRevD.5.1916,PhysRevD.7.2825,PhysRevD.7.2838,PhysRevD.8.1645,PhysRevD.15.377,PhysRevD.24.327,PhysRevD.31.1236,PhysRevD.31.1252,PhysRevD.34.1025},
all of which are shown to approximate the form of the Maxwell
and Lorentz-Dirac equations given above.
We will not consider these other variations of our theory here except to
remark that resulting differences in higher order corrections
may yield slightly different physical predictions, and thus may potentially
be constrained by observation.

Now we note that the above equations deviate from conventional 
classical electrodynamics when $k^2$ is non-negligible,
as Eqs.~(\ref{maxwell}-\ref{maxwelllast}) imply that the electric field of a static point charge is 
$q(r^{-2}-k^2)\hat{r}$.  
At distances on the scale of $\ell$ from a source current, then, there is 
significant violation of Coulomb's law.  
Therefore some of the same tests used to bound the length scale of the 
well-known Coulomb-violating Yukawa potential can be used to constrain $\ell$.
In addition to laboratory measurements of the electrostatic force,
satellite measurements of the Earth's magnetic field indicate $\ell\gtrsim 10^8\m$ \cite{RevModPhys.43.277},
Pioneer 10's measurement of Jupiter's magnetic field indicate $\ell\gtrsim 10^9\m$ \cite{1975PhRvL..35.1402D}, and measurements of solar wind electrodynamics indicate $\ell\gtrsim 10^{11}\m$ \cite{0741-3335-49-12B-S40}.
(See \cite{Goldhaber:2008xy} for a review of such Coulomb tests.)
We will assume this latter bound. 

On the scale of $\ell$, our theory exhibits two remarkable features that are 
relevant to our current purposes.  The first is that it allows 
equilibrium configurations of charges not possible in Einstein-Maxwell theory.
For example, two identically
charged particles at rest and separated by $2\ell$ reside in the zeros of each
other's electrostatic fields.  Such an equilibrium may even be stable,
as in that example.  In Einstein-Maxwell theory
charged particles cannot be in stable equilibrium because Laplace's equation
disallows the needed extrema in the potential, a fact known as Earnshaw's
Theorem.  But Einstein's nonsymmetric field equations are
third order and thus not subject to Earnshaw's Theorem.

A second notable feature is the factor of 2 multiplying the electromagnetic
field tensor $F^D_{\rho\sigma}$ in Eq.~(\ref{F2gammastar}), which determines its coupling
to the gravitational field, which is absent in the Lorentz force equation.
This asymmetry between the electromagnetic field's coupling to charged
particles vs its coupling to the gravitational field has an unexpected result:
the electromagnetic field $F^L+F^D$ can cancel with itself and thus vanish
in the Lorentz force equation, while $F^L+2F^D$ remains nonzero and still
acts as a source for the gravitational field.  In this sense,
the electromagnetic field can become ``dark".  In this case
\begin{eqnarray}
\gamma^{[\mu\nu]} &=& c^{-2}G^{1/2}k^{-1}\epsilon^{\mu\nu\rho\sigma}(F^L_{\rho\sigma}+2F^D_{\rho\sigma})\nonumber\\
&=& c^{-2}G^{1/2}k^{-1}\epsilon^{\mu\nu\rho\sigma}F^D_{\rho\sigma}.
\end{eqnarray}

\section{``Dark matter" solutions}

\subsection{Electromagnetic equilibrium}

For charged matter to be in equilibrium, the force density must vanish:
\begin{equation}
\label{eq:equilibrium}
J^\nu(F^L_{\mu\nu}+F^D_{\mu\nu})=0.
\end{equation}
Now following \cite{johnson} we derive the condition for 
electrostatic equilibrium.
Neglecting magnetic fields,
Eq.~(\ref{eq:equilibrium}) becomes
\begin{equation}
\label{eq:electrostaticequilibrium}
\rho(\mathbf{E}_L+\mathbf{E}_D)=0,
\end{equation}
where
\begin{eqnarray}
\nabla\cdot\mathbf{E}_L &=& 4\pi\rho,\\
\nabla\cdot\mathbf{E}_D &=& -k^2\varphi,
\end{eqnarray}
and the Maxwell potential $\varphi$ is defined such that $\mathbf{E}_L=-\nabla\varphi$.
Wherever the charge density is nonvanishing the equilibrium
condition takes the form of the Helmholtz equation:
\begin{equation}
\label{helmholtz}
\nabla^2\varphi+k^2\varphi=0.
\end{equation}
If we assume the charge density has unbounded support but vanishes at infinity 
then the general solution is
\begin{equation}
\varphi(r,\theta,\phi)=\frac{4\pi\rho_{0}}{k^2}\sum_{l=0}^\infty\sum_{m=-l}^l [a_{l m}j_l(kr)+b_{l m}y_l(kr)]\Y_{l m}(\theta,\phi)
\end{equation}
where $\rho_{0}$ is a scaling constant with units of charge density.
Note the above equations imply $\rho=\frac{k^2}{4\pi}\varphi$.
Solutions also exist for a charge density with bounded support, which might be
more physical, but for now we will assume this simpler case.

Of course in a more realistic model the charges will be in motion and thus 
generate a magnetic field.  We could for example allow magnetic fields without 
altering the charge distribution by requiring that the source currents respect 
the symmetry of the distribution (as in a rotating, axially symmetric case).  
In that case, since the distribution is such that its electrostatic 
contribution to the Lorentz force cancels, it is possible that there may be 
some cancellation of the magnetic contribution as well (since it is similarly 
modified on large scales [Eq.~(\ref{lorentz}]).  Alternatively, we could attempt to find 
more general solutions to the equilibrium condition [Eq.~(\ref{eq:equilibrium})] by relaxing the 
assumption of vanishing 3-current and magnetic field.  We might therefore 
obtain a magnetic field that is dark, or nearly dark.   But as this would 
require a considerably more complicated model, and the electrostatic case of 
the present work suffices as a ``proof of principle", we relegate the case of 
nonvanishing magnetic field to a future work.

The above solution is stable to small, local perturbations for the simple
reason that any small displaced charge will create an oppositely charged ``hole"
in the ambient equilibrium distribution, to which it will be attracted
provided the displacement is much smaller than $2\ell$.
But it is not clear under what conditions the above solution is stable to larger
perturbations.  To thoroughly address this question, it may be necessary
to consider more realistic solutions with nonzero magnetic fields.

If we wish to apply this solution to a galaxy there are two additional caveats.
The first is that $\rho$ must be averaged over lengths much larger than
the Debye length.  We are interested in net charge density that survives when
averaged over scales closer to $\ell$.  We expect this to be no greater than
the more localized (sub-Debye) plasma density, and perhaps smaller 
due to Debye shielding of the latter.

Second, if we assume that $r=0$ corresponds to the galactic center, 
then observation suggests we further assume an uncharged black hole there. 
(For relaxation of this assumption, see the Appendix.)
Thus the charge density must be negligible near the origin.
We therefore take as a boundary condition
that $\rho(0)=0$, which implies
\begin{equation}
\varphi(r,\theta,\phi)=\frac{4\pi}{k^2}\sum_{l=1}^\infty\sum_{m=-l}^l a_{l m}j_l(kr)Y_{l m}(\theta,\phi).
\end{equation}

We will be interested in the resulting gravitational field.
As mentioned, when the electromagnetic fields are in equilibrium
the gravitational field couples to $F^L_{\mu\nu}+2F^D_{\mu\nu}=F^D_{\mu\nu}$.
In the static case, the time-time component of the gravitational field
equation reduces to
\begin{equation}
\nabla^2\gamma_{00}=-t_{00},
\end{equation}
where
\begin{equation}
t_{00} = -\frac{1}{4}\gamma_{[\rho\sigma],\kappa}\gamma^{[\rho\sigma],\kappa}
-\frac{1}{2}\gamma_{[\rho\sigma],\kappa}\gamma^{[\rho\kappa],\sigma}
+\frac{1}{2}\gamma^{[\rho\sigma]}\nabla^2\gamma_{[\rho\sigma]},
\end{equation}
and for electrostatic equilibrium,
\begin{equation}
\gamma^{[\mu\nu]}=2c^{-2}G^{1/2}k^{-1}\epsilon^{\mu\nu0\sigma}\partial_\sigma\varphi.
\end{equation}

\subsection{Fields at large radius}
\label{sec:largeradius}

The asymptotic form of the $l$th spherical Bessel function is \cite[Eq.~(10.52.3)]{NIST:DLMF}
\begin{equation}
\label{asymptotic}
j_l(kr)=\frac{\sin(kr-l\pi/2)}{kr}+O(r^{-2})
\end{equation}
In \cite{Heitman2015347} this is shown to be a valid approximation for $kr>l$.
With this approximation the potential becomes
\begin{widetext}
\begin{eqnarray}
\varphi(r,\theta,\phi) &=& \frac{4\pi}{k^2}\sum_{l>0}\sum_{m=-l}^l a_{l m}\frac{\sin(kr-l\frac{\pi}{2})}{kr}Y_{l m}(\theta,\phi)+O(r^{-2})\nonumber\\
&=& \frac{4\pi}{k^2}[\Bj_0(kr)\sum_{\substack{l>0\\\text{ even}}}\sum_{m=-l}^l a'_{l m}\Y_{l m}(\theta,\phi)+\By_0(kr)\sum_{\substack{l>0\\\text{ odd}}}\sum_{m=-l}^l b'_{l m}\Y_{l m}(\theta,\phi)]+O(r^{-2}),
\end{eqnarray}
where $a'_{l m}\equiv(-1)^{l/2}a_{l m}$ and 
$b'_{l m}\equiv(-1)^{(l-1)/2}a_{l m}$.
The spherical harmonics form a complete set of functions on the sphere,
which implies that the sum on the left converges to an even function
(with respect to $\theta=\pi/2$, and with vanishing spherical integral since we are missing the constant
spherical harmonic $\Y_{00}$),
while the sum on the right converges to an odd function (with respect to
$\theta=\pi/2$).
Hence
\begin{equation}
\label{eq:evenodd}
\varphi(r,\theta,\phi)= \frac{4\pi}{k^2}[\Bj_0(kr)f(\theta,\phi)+\By_0(kr)g(\theta,\phi)]+O(r^{-2}),
\end{equation}
where $f(\theta,\phi)$ is any even square-integrable function on the sphere such
that $\int fd\Omega=0$, and $g(\theta,\phi)$ is any odd square integrable function on the
sphere.  

Then the electric field is
\begin{equation}
\gamma^{[\mu\nu]}=-2G^{1/2}\frac{4\pi}{c^2k^2}\epsilon^{\mu\nu 0\sigma}
[\Bj_1(kr)f(\theta,\phi)+\By_1(kr)g(\theta,\phi)]x_\sigma r^{-1}+O(r^{-2}).
\end{equation}
Then using
\begin{eqnarray}
\partial_i[\Bj_1(kr)n_j] &=& k\left\{ \frac{1}{3}[\Bj_0(kr)+\Bj_2(kr)]\delta_{ij}-\Bj_2(kr)\frac{x_ix_j}{r^2}\right\},\\
\partial_i[\By_1(kr)n_j] &=& k\left\{ \frac{1}{3}[\By_0(kr)+\By_2(kr)]\delta_{ij}-\By_2(kr)\frac{x_ix_j}{r^2}\right\},
\end{eqnarray}
the effective gravitational source is
\begin{eqnarray}
t_{00} &=& G\left(\frac{4\pi \rho_{0}}{c^2k}\right)^2\left\{\left[-\frac{1}{6}(\Bj_0(kr))^2-(\Bj_1(kr))^2+\frac{2}{3}(\Bj_2(kr))^2\right]f^2(\theta,\phi)\right.\nonumber\\
&& + \left[-\frac{1}{3}\Bj_0(kr)\By_0(kr)-2\Bj_1(kr)\By_1(kr)+\frac{4}{3}\Bj_2(kr)\By_2(kr)\right]f(\theta,\phi)g(\theta,\phi)\nonumber\\
&& + \left.\left[-\frac{1}{6}(\By_0(kr))^2-(\By_1(kr))^2+\frac{2}{3}(\By_2(kr))^2\right]g^2(\theta,\phi)\right\}+O(r^{-3}).
\label{eq:asymptoticsource}
\end{eqnarray}
Now we seek a solution with vanishing derivative at spatial infinity.
We obtain, up to an integration constant that will not affect the acceleration, 
\begin{eqnarray}
\gamma_{00} &=& G\left(\frac{2\pi \rho_{0}}{c^2k^2}\right)^2\left\{[\ln(2kr)-\Ci(2kr)+\Bj_0(2kr)-2(\Bj_1(kr))^2]f^2(\theta,\phi)\right.\nonumber\\
&&+[-2\Si(2kr)-\By_0(2kr)-4\Bj_1(kr)\By_1(kr)]f(\theta,\phi)g(\theta,\phi)\nonumber\\
&&+\left.[\ln(2kr)+\Ci(2kr)-\Bj_0(2kr)-2(\By_1(kr))^2]g^2(\theta,\phi)\right\}+O(r^{-1}).
\end{eqnarray}

To calculate the resulting acceleration of a test mass (in a Newtonian approximation),
first we note that to linear order the densitized metric perturbation equals the
trace-reversed metric perturbation,
\begin{equation}
\label{eq:tracereversed}
\gamma_{(\mu\nu)}=-h_{\mu\nu}+\frac{1}{2}\eta_{\mu\nu}\eta^{\rho\sigma}h_{\rho\sigma},
\end{equation}
where $h_{\mu\nu}\equiv g_{(\mu\nu)}-\eta_{\mu\nu}$.
In the harmonic gauge, the trace-reversed metric perturbation is in turn 
proportional to the Newtonian potential.
Following Wald \cite[\S 4.4]{Wald:106274}, 
we find for the radial acceleration
\begin{eqnarray}
a_r &=& \frac{1}{4}c^2\partial_r\left(h_{00}-\frac{1}{2}\eta_{00}\eta^{\rho\sigma}h_{\rho\sigma}\right)\nonumber\\
&=& -\frac{1}{4}c^2\partial_r\gamma_{00}\nonumber\\
&=& -G\left(\frac{\pi \rho_{0}}{ck^2}\right)^2\left\{[1-\Bj_0(2kr)-4\sin(kr)\Bj_1(kr)+8\Bj_1^2(kr)]f^2(\theta,\phi)\right.\nonumber\\
&&+[3\By_0(2kr)-8\Bj_0(kr)\By_0(kr)+16\Bj_1(kr)\By_1(kr)]f(\theta,\phi)g(\theta,\phi)\nonumber\\
&&+\left.[1+\Bj_0(2kr)+4\cos(x)\By_1(kr)+8(\By_1(kr))^2]g^2(\theta,\phi)\right\}r^{-1}+O(r^{-2})\nonumber\\
&=& -G\left(\frac{\pi \rho_{0}}{ck^2}\right)^2[f^2(\theta,\phi)+g^2(\theta,\phi)]r^{-1}+O(r^{-2}).
\end{eqnarray}
\end{widetext}

The angular functions $f$ and $g$ must be determined empirically.
Aside from the constraint that $f$ must average to zero over a sphere,
there is considerable freedom to choose these functions.
For example 
they can be chosen such that isosurfaces of $f^2+g^2$ are approximately ellipsoidal.
Gravitational lensing observations of the Milky Way's dark matter halo \cite{2010AAS...21532103L}, as well as the success of the pseudo-isothermal elliptic (PIE) mass 
distribution at modeling many other galaxies and clusters \cite{1993ApJ...417..450K}, 
indicate that such symmetry is relevant. 
On the other hand, if we choose
\begin{eqnarray}
\label{eq:sin}
f(\theta,\phi) &=& \sin(n\theta),\\
g(\theta,\phi) &=& \cos(n\theta),
\label{eq:cos}
\end{eqnarray}
where $n>1$ is odd,
then we achieve spherical symmetry.
For simplicity, and because it seems a good approximation for
many purposes, we will focus on this case.

The scaling factor $\rho_{0}$ 
must also be determined empirically for each galaxy.  
Fitting each galactic rotation curve requires that
\begin{equation}
G\left(\frac{\pi \rho_{0}}{ck^2}\right)^2=v^2,
\end{equation}
where $v$ is the constant speed that characterizes dark-matter-dominated orbital dynamics.
For example, using the observed Milky Way value of $v\approx 240 \km/\s$ \cite{doi:10.1093/pasj/64.6.136,0004-637X-783-2-130,doi:10.1093/mnras/stw2096} we obtain the formula
\begin{equation}
\label{eq:density}
\rho_0\approx\left(\frac{9.6\times 10^9\km}{\ell}\right)^2\e/\cm^{3}.
\end{equation}

Note that Eq.~(\ref{eq:density}) may result in a density considerably smaller 
than 
that of plasmas that have been observed in and around the galaxy.  For example, 
the warm ionized medium has an ion density greater than $10^{-2}\e/\cm^3$, 
while the galactic halo has been observed to have an ion density greater than 
$10^{-4}\e/\cm^3$ \cite{2001RvMP...73.1031F}.  By contrast, if we consider the region where dark matter 
dominates the Milky Way, $r\gtrsim 8\kpc$, so that the asymptotic expression 
Eq.~(\ref{eq:evenodd}) is approximately valid, and using our constraint that 
$\ell\gtrsim 10^{11}\m$, we obtain $\rho<10^{-5}\e/\cm^3$.  And $\rho$ may be 
orders of magnitude smaller still, since $\ell$ may be correspondingly larger 
(see Sec.~VI.C and VI.D).  One implication is that, in a more realistic
model in which the charge distribution consists of discrete particles, any 
observable electromagnetic effects due to the resulting imperfect cancellation 
of fields on small scales may be swamped by that of more conventional plasmas.

Now revisiting the gravitational field, recall that Eq.~(\ref{eq:tracereversed}) and Wald \cite[\S 4.4]{Wald:106274} imply a Newtonian potential approximately equal to $\frac{1}{4}c^2\gamma_{00}$.  This implies an effective mass density of 
\begin{equation}
\frac{c^2}{16\pi G}\nabla^2\gamma_{00}=-\frac{c^2}{16\pi G}t_{00}.  
\end{equation}\\
Then using the effective source of Eq.~(\ref{eq:asymptoticsource}) with the 
asymptotic Bessel forms of Eq.~(\ref{asymptotic}) and the angular functions of Eqs.~(\ref{eq:sin}, \ref{eq:cos}), and assuming a sufficiently large $n$, we obtain to good approximation an angle-averaged mass density of $\frac{v^2}{4\pi Gr^2}$.  This expression is also the 
asymptotic limit of the pseudo-isothermal mass density profile for dark matter
\cite{Gunn:1972sv,1991MNRAS.249..523B}. The pseudo-isothermal model has in turn been found to be a good 
fit for the rotation curves of many galaxies \cite{deBlok:2008wp}, as well as for the mass
density of some galaxy clusters as determined by gravitational lensing and x-ray temperatures \cite{1996IAUS..173...49R,Revaz:2006zv,Bonamigo:2017idz}.

The pseudo-isothermal model for dark matter, however, is not always the best
fit.  We have already mentioned its generalization, the PIE model for 
ellipsoidal symmetry,
which might be approximated by different choices for the angular
functions $f$ and $g$ above.  We might also consider truncating the charge 
distribution $\rho$ so that it has bounded support.  In this case, 
we expect solutions of Eq.~(\ref{eq:electrostaticequilibrium}) to still approximate those given above, 
within the support of $\rho$.  But the fields and thus effective mass density
will fall off more rapidly outside of that support, with a smooth transition
in between.  Such behavior also characterizes a further generalization
of the pseudo-isothermal model, called the ``truncated" or ``dual" PIE model, 
which very successfully fits a wide range of lensing and x-ray data \cite{1993ApJ...417..450K,Limousin:2004mj,Eliasdottir:2007md,0004-637X-800-1-38,1742-6596-689-1-012005,Bonamigo:2017idz}.
Further exploration of our dark-matter-like solution space might be worthwhile,
therefore.
We will for now, however, content ourselves to 
study our pseudo-isothermal-like solution, the structure of which
is already quite rich.\\

\subsection{Potential at all radii}

\subsubsection{Convergence}
The asymptotically spherical solution found in the last section is,
in its exact form,
\begin{widetext}
\begin{equation}
\varphi_{(n)}(r,\theta)=\frac{4\pi\rho_{0}}{k^2}\sum_{l=1}^\infty a_l(n)\Bj_l(kr)\LP_l(\cos(\theta)),
\end{equation}
where
\begin{equation}
a_l(n)=\frac{1}{2}(2l+1)\left\{\begin{array}{ll} (-1)^{(l-1)/2}\int_0^\pi\LP_l(\cos(\theta))\cos(n\theta)\sin(\theta)d\theta,& l\text{ odd}\\
(-1)^{l/2}\int_0^\pi\LP_l(\cos\theta)\sin(n\theta)\sin(\theta)d\theta,& l\text{ even}\end{array}\right.
\end{equation}
\end{widetext}
We have established, by construction using the completeness of the Legendre polynomials
along with the square-integrability of $\cos(n\theta)$ and $\sin(n\theta)$,
that asymptotically the above converges to a linear combination of 
those functions.  But this is not generally the case at other radii,
because the Bessel functions may not cancel the $(-1)$ factors.
Our next task, then, is to verify that the series still converges at all radii.

Since the issue is the sign on the terms, it will suffice to show
that the series converges absolutely.  First note that since
$\cos(n\theta)$ equals an $n$th order (Chebyshev) polynomial in $\cos(\theta)$,
only a finite number of Legendre polynomials are required in its expansion,
and thus there are only a finite number of nonvanishing $a_l(n)$ with odd $l$.
So we need only show the absolute convergence of the even terms.

To that end we have, using a Chebyshev identity and the binomial theorem,
\begin{eqnarray}
\sin(n\theta) &=& \sin(\theta)U_{n-1}(\cos(\theta))\nonumber\\
&=& (1-\cos^2(\theta))^{1/2}U_{n-1}(\cos(\theta))\nonumber\\
&=& \sum_{i=0}^\infty(-1)^i\left(\begin{array}{c} 1/2 \\ i \end{array}\right)\cos^{2i}(\theta)U_{n-1}(\cos(\theta))\nonumber\\
&=& \sum_{j=0}^\infty c_j\cos^j(\theta),  
\end{eqnarray}
where $U_{n-1}$ is the $(n-1)$th order Chebyshev polynomial of the second kind,
and $\{c_j\}$ are real coefficients defined by the expansion above.
Since the binomial expansion converges absolutely,
and $U_{n-1}$ equals only a finite number of powers of $\cos(\theta)$,
the entire series converges absolutely. 
Since, then, $\sum_j|c_j\cos^j(\theta)|$ converges on $[0,\pi]$,
it is square-integrable 
and thus admits a unique expansion in Legendre polynomials.
This expansion can be calculated by first expressing each power of $\cos(\theta)$ in Legendre
polynomials, which have only positive coefficients in this case.
It follows that $\sum_{l>0,\text{ even}}|a_l(n)\LP_l(\cos(\theta))|=\sum_j|c_j\cos^j(\theta)|<\infty$.
Since furthermore the Bessel functions are bounded by 1, 
they do not affect the absolute convergence.
We conclude that $\varphi_{(n)}$
is a valid solution at all radii.

\subsubsection{Core radius}
\label{sec:coreradius}

As mentioned, the gravitational field of this solution bears some resemblance to that of
the pseudo-isothermal model of
dark matter \cite{Gunn:1972sv,1991MNRAS.249..523B}, 
in that at large radius it results in a flat rotation curve
(as shown in Section~\ref{sec:largeradius}), while at small radius it is
negligible (as will be shown in the next section).
And like the pseudo-isothermal model
it has a free parameter to set the effective mass scale and thus 
fit the flat rotation curve of each galaxy.
The pseudo-isothermal model also has a free parameter,
known as the ``core radius",
which sets the length scale at which dark matter transitions from 
having subdominant effect to
determining flat
rotation dynamics. 
We claim that $n\ell$ plays a similar role,
where $n$ is the free, odd integer parameter in $\varphi_{(n)}$ above.

\begin{figure}
\includegraphics[scale=0.4]{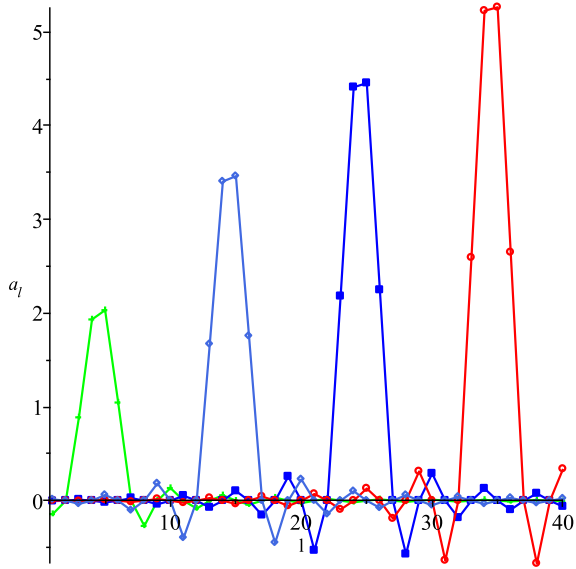}
\caption{Coeffecients $a_l(5)$ (crosses), $-a_l(15)$ (diamonds), $a_l(25)$ (squares), and $-a_l(35)$ (circles) vs $l$.  This shows that $|a_l(n)|$ is largest
when $l\approx n$.}
\label{fig:coeffs}
\end{figure}

To justify this claim, first note that from various properties
of Legendre and Chebyshev polynomials \cite{abramowitz+stegun,Legendre,Chebyshev,MultipleAngle} it follows that
\begin{widetext}
\begin{enumerate}[label=(\roman{*})]
\item If $l$ is even and $l<n-1$ then $a_l(n)=0$. 
\item If $l$ is odd and $l\leq n$ then
\begin{equation}
a_l(n)=\frac{(-1)^{(l-3)/2}2^{2l-1}n(2l+1)
[(n+l-2)/2]![(n+l)/2]!(n-l)!}{(n-l-1)[(n-l)/2)!]^2(n+l+1)!}.
\end{equation}
\item If $l$ is odd and $l>n$ then $a_l(n)=0$.
\end{enumerate}
\end{widetext}
And from (i) and (ii) above it follows that 
$|a_n(n)|\gg|a_l(n)|$ for $l<n-1$.  
This is shown numerically in Fig.~\ref{fig:coeffs}, where we also observe that
$|a_n(n)|\gg|a_l(n)|$ for $l>n+1$.

\begin{figure*}
\includegraphics[scale=0.6,angle=-90]{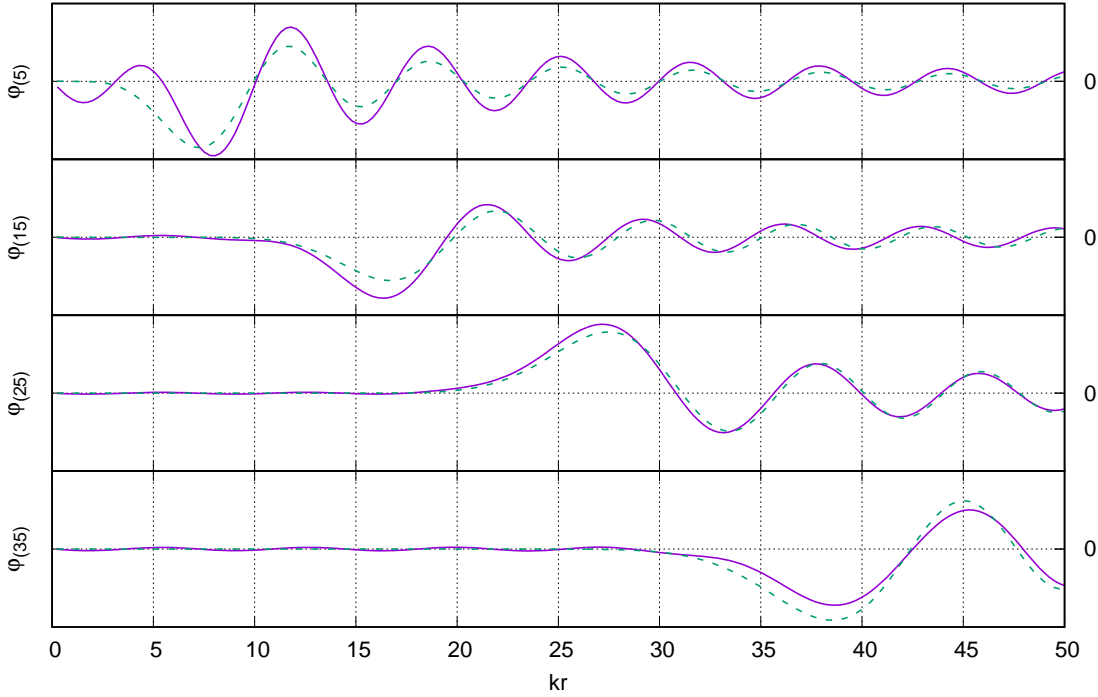}
\caption{
The potential $\varphi_{(n)}(r,\theta=\pi/6)$ for $n=5$, $n=15$, $n=25$, and $n=35$ (as calculated numerically)
plotted together with an approximation (dashed line) using only 
the $l=n-1$, $l=n$, and $l=n+1$ terms:
$
\frac{4\pi\rho_{0}}{k^2}\sum_{l=n-1}^{n+1} a_l(n)\Bj_l(kr)\LP_l(\cos(\theta)). 
$
This shows that the potential is dominated by terms around $l\approx n$.
This is further made clear by the global extremum apparent near $kr\approx n+2$,
which approximates the position of the global extremum of $\Bj_n(kr)$.
And, like $\Bj_n(kr)$, the potential is relatively negligible for $kr\lesssim n$.
}
\label{fig:potential}
\end{figure*}

From the above results, we expect
the potential $\varphi_{(n)}$ may be dominated by the
$n$th Bessel function (and its neighbors the $(n-1)$th and $(n+1)$th
Bessel functions).
Such is clearly evident in the $\theta=\pi/2$ plane,
since the odd polynomials in $\cos(\theta)$ vanish and the first nonvanishing
term is the $(n-1)$th term.
The situation is more complicated away from this plane, but
the numerical results exemplified by Fig.~\ref{fig:potential} indicate that our expectations
are well founded for $kr>1$ (and for $kr<1$ we will see the potential is negligible).

In particular, Fig.~\ref{fig:potential} illustrates that the global extremum of $\varphi_{(n)}(r,\theta)$ is near $kr\approx n+2$ (we have numerically found such to the be case
for all values of $\theta$).  According to Abramowitz and Stegun, 
for large $n$ the location of the global extremum of $\Bj_n(kr)$ 
(which we shall call $r_n$)
is \cite{abramowitz+stegun}:
\begin{widetext}
\begin{equation}
kr_n\approx n+\frac{1}{2}+0.8086165\left(n+\frac{1}{2}\right)^{\frac{1}{3}}-0.236680\left(n+\frac{1}{2}\right)^{-\frac{1}{3}}-0.20736\left(n+\frac{1}{2}\right)^{-1}+0.0233\left(n+\frac{1}{2}\right)^{-\frac{5}{3}}.
\end{equation}
\end{widetext}
In \cite{Kosma:13} it is further observed that this is a good approximation for all $n\geq 5$
(for which its relative error is $<1\%$).
This formula, in turn, is approximated by $kr_n\approx n+2$ 
for all $n\geq 5$ to within a relative error of $5\%$. 

More to the point, Fig.~\ref{fig:potential} demonstrates that $\varphi_{(n)}$ further shares with $\Bj_n(kr)$ the property that it is relatively small for $kr<n$, 
or $r<n\ell$, at least for $n>5$.  
The derivatives of $\varphi_{(n)}$ are also relatively small for $r<n\ell$,
and thus so is the resulting gravitational field.
This supports our designation of $n\ell$ as the core radius.

We conclude this section with the following observation.
As noted in Section~(\ref{sec:largeradius}), the asymptotic form of the $l$th Bessel 
function Eq.~(\ref{asymptotic}) is valid for $kr>l$. 
If as we have argued $\varphi_{(n)}$ is dominated by the $n$th Bessel function,
then the analysis of the previous section implies that 
the rotation curve can be assumed flat for $r\gg n\ell$.

\subsection{Fields at small radius}

We consider again the Milky Way, for a concrete example.
In order to be a viable model of dark matter in the Milky Way, 
our solution must also
be consistent with observations of Sagittarius A* (Sgr A*).  
First, its gravitational field should be negligible in the region around
Sgr A* where Keplerian stellar orbits have been observed.  
And second, charge density of our equilibrium distribution should 
be neglible in the immediate vicinity of the central black hole.

Recall that we assume that $\ell\gtrsim 10^8\km >10^7\km$, the Schwarzschild radius of Sgr A*.
We will consider a region where $kr<1$ ($10^7\km<r<\sqrt{2}\ell$), such that $\Bj_l(kr)$ is
well approximated by the lowest order terms of a Taylor expansion in $kr$. 
In this region the lower order Bessel functions can be expected to dominate since \cite[Eq.~(10.52.1)]{NIST:DLMF}
\begin{equation} 
\Bj_l(kr)=(kr)^l/(2l+1)!!+O((kr)^{l+2}).
\end{equation}
It is also useful to note that the bound
\begin{equation}
|\Bj_l(kr)|\leq(kr)^l/(2l+1)!!
\end{equation}
is valid everywhere \cite[Eq.~(10.14.4)]{NIST:DLMF}.

\subsubsection{Gravitational field}

To third order in $(kr)$ the potential can be expressed
\begin{widetext}
\begin{equation}
\varphi_{(n)} = \frac{4\pi\rho_{0}}{k^2}\left[a_1(n)\Bj_1(kr)\LP_1(\cos(\theta))+a_2(n)\Bj_2(kr)\LP_2(\cos(\theta))+a_3(n)\Bj_3(kr)\LP_3(\cos(\theta))\right]+O(r^4),\\
\end{equation}
where, assuming $n>3$,
\begin{eqnarray}
a_1(n) &=& -\frac{3}{n^2-4},\\
a_2(n) &=& 0,\\
a_3(n) &=& \frac{7(n^2-1)}{n^4-20n^2+64}.
\end{eqnarray}
Taylor expanding the Legendre polynomials explicitly,
\begin{eqnarray}
\varphi_{(n)} &=& \frac{4\pi\rho_{0}}{k^2}\left[a_1(n)\left(\frac{1}{3}(kr)-\frac{1}{30}(kr)^3\right)\cos(\theta)+a_3(n)\frac{1}{105}(kr)^3\left(\frac{5}{2}\cos^3(\theta)-\frac{3}{2}\cos(\theta)\right)\right]+O(r^4)\nonumber\\
&=& \frac{4\pi\rho_{0}}{k^2}\left[a_1(n)\left(\frac{1}{3}kz-\frac{1}{30}k^3r^2z\right)+a_3(n)\frac{1}{105}k^3\left(\frac{5}{2}z^3-\frac{3}{2}r^2z\right)\right]+O(r^4).
\end{eqnarray}
To calculate the leading order of $t_{00}$, it will prove sufficient to
calculate the $z$-component of $\mathbf{E}^D$:
\begin{equation}
E^D_z=\frac{4\pi\rho_{0}}{k}\left[a_1(n)\frac{1}{3}\left(1-\frac{1}{5}k^2z^2-\frac{1}{10}k^2r^2\right)+a_3(n)\frac{1}{70}k^2\left(3z^2-r^2\right)\right]+O(r^3).
\end{equation}
Then
\begin{eqnarray}
t_{00} &=& \frac{1}{2}\gamma^{[\rho\sigma]}\nabla^2\gamma_{[\rho\sigma]}+O(r^1)\nonumber\\
&=& G\left(\frac{4\pi\rho_{0}}{k^2c^2}\right)^2\left\{\frac{1}{3}a_1(n)k^2\nabla^2\left[-\frac{1}{3}a_1(n)\left(\frac{1}{5}z^2+\frac{1}{10}r^2\right)+\frac{1}{70}a_3(n)(3z^2-r^2)\right]\right\}+O(r^1).
\end{eqnarray}
So
\begin{equation}
\gamma_{00}=G\left(\frac{4\pi\rho_{0}}{k^2c^2}\right)^2\left\{\frac{1}{3}a_1(n)k^2\left[\frac{1}{3}a_1(n)\left(\frac{1}{5}z^2+\frac{1}{10}r^2\right)-\frac{1}{70}a_3(n)(3z^2-r^2)\right]\right\}+O(r^3).
\end{equation}
Therefore
\begin{eqnarray}
a_r &=& -\frac{1}{4}c^2\partial_r\gamma_{00}\nonumber\\
&=& -4G\left(\frac{\pi\rho_{0}}{k^2c}\right)^2\left\{\frac{1}{3}a_1(n)k^2\left[\frac{1}{3}a_1(n)\left(\frac{2}{5}r\cos^2(\theta)+\frac{1}{5}r\right)-\frac{1}{35}a_3(n)(3r\cos^2(\theta)-r)\right]\right\}+O(r^2).
\end{eqnarray}
\end{widetext}
Then using $a_1(n)\approx -3n^{-2}$ and $a_3(n)\approx 7n^{-2}$, 
and recalling $G\left(\frac{\pi\rho_{0}}{k^2c}\right)=v^2$,
we obtain
\begin{equation}
a_r\approx -\frac{2v^2r}{n^4\ell^2}\cos^2(\theta).
\end{equation}

According to recent observations of the star cluster around Sgr A*,
a radial acceleration of 
\begin{equation}
-\frac{G(4\times 10^6 M_\odot)}{r^2} \approx -\frac{5\times 10^{17}\km^3s^{-2}}{r^2} 
\end{equation}
due to the 
 central black hole appears to dominate stellar orbits
up to about $10^{13}\km$ \cite{refId0}.  
In order to apply our small-radius ($r<\ell$) approximation to the region 
$r<10^{13}\km$ we will assume that $\ell\gtrsim 10^{13}\km$.
We then find the magnitude of the acceleration $a_r$ is less than that
due to the black hole for $r<10^{13}\km$ provided $n^2\ell\gtrsim 10^{13}\km$.
This opens up the possibility that $\ell<10^{13}\km$, but verification would require
a solution that is more accurate at intermediate ($r>\ell$) radii.
For now we note that $\ell=10^{13}\km$ suffices.
With that value, and assuming $n\geq 5$, the acceleration
$|a_r|$ is less than a millionth that due to the black hole at
$r=3000$~AU, just inside the region where the cleanest Keplerian
orbits have been observed \cite{Ghez:2003qj,0004-637X-628-1-246}. 

\subsubsection{Charge}

We turn our attention now to the charge density in the immediate vicinity
of the central black hole.
The danger here is not that the black hole will acquire charge, as our solution
is such that the net charge within any centered, spherical volume is zero.  But
within that volume, there is separation of charge which generates an electric
field that contributes to the equilibrium of the surrounding distribution.
If the black hole neutralizes these separated charges by unseparating 
them, it could threaten that equilibrium.  We will show that the
charge so endangered is negligible.

Because we are concerned with charge carriers
regardless of sign, we will consider the absolute value of the charge density:
\begin{eqnarray}
|\rho| &=& |\rho_0a_1(n)\Bj_1(kr)\cos(\theta)|+O(r^3)\nonumber\\
&=& \frac{1}{3}k|\rho_0a_1(n)r\cos(\theta)|+O(r^3)
\end{eqnarray}
Volume integration within a radius $R$ yields
\begin{eqnarray}
Q &=& \int_{r<R}|\rho|d^3r\nonumber\\
&=& \frac{2\pi}{3}|\rho_0a_1(n)|k\int_0^R\int_0^\pi r^3|\cos(\theta)|\sin(\theta)d\theta+O(r^6)\nonumber\\
&=& \frac{\pi}{2(n^2-4)}|\rho_0|kR^4+O(r^6)
\end{eqnarray}
Now we assume $\ell=10^{13}\km$ as suggested above. 
Regarding $n$, if we further assume that $n\ell$ is comparable to the
core radius of the pseudo-isothermal model as suggest in Sec.~\ref{sec:coreradius},
then for the Milky Way $n\ell\gtrsim 1\kpc$ \cite{MNR:MNR9367} and thus $n\approx 3000$.
Then using the Sgr A* Schwarzschild radius of $\sim 10^7\km$ for $R$ 
and recalling $|\rho_0|\approx e(10^{10}\km/\ell)^2\cm^{-3}$ from the fit to the Milky Way's
flat rotation curve, we find that $Q\approx 0.02$~C.
This seems negligible as, for comparison, the net charge of the Sun is 77~C \cite{2001A&A...372..913N}.

\section{Discussion}

Einstein's nonsymmetric field theory 
allows non-neutral plasma to be in electromagnetic equilibrium on the scale of a universal 
length constant.  
We have shown, in the context of this theory, that the gravitational field generated by the electrostatic field of a 
certain family of such equilibrium solutions resembles that of the pseudo-isothermal 
model for dark matter.  As such it can be fit to the mass density profile of some 
galaxies and clusters, as determined by velocity, lensing, and x-ray data.  Further we 
have argued that generalizations of these solutions may approximate generalizations of 
the pseudo-isothermal model - specifically the truncated pseudo-isothermal elliptic 
model.  This is one of the best phenomenological models for dark matter, fitting a wide 
range of galaxies and clusters, and thus gives reason to hope for similar success from 
these solutions.

However, some significant challenges remain in proving the physical relevance 
of this theory and of these solutions in particular.  
One is the question of the compatibility of Einstein's nonsymmetric field theory with 
inflationary cosmology.  This may need to be addressed in order to understand its 
bearing on the cosmic microwave background power spectrum, baryon acoustic oscillations, and structure 
formation, all of which are significantly affected by dark matter according to the 
Lambda-CDM paradigm.  But these signatures of the early universe are well beyond the 
purview of the classical, post-Minkowskian methods used in the present work.

Other open questions include the value of the universal length constant $\ell$,
for which our suggestion is far from definitive, and which is not well constrained
by the data considered in this paper.  And perhaps most crucially, the stability
of our solutions to large perturbations needs to be established, up to and including
the Bullet Cluster collision.  The intent of this paper is to make the case that
these inquiries are worthwhile, and to open up a new approach to an old problem.

\appendix
\section*{Appendix: Spherically symmetric solution}

In this appendix we consider a spherically symmetric solution originally found 
by Johnson \cite{johnson}.
The simplest nontrivial solution to Eq.~(\ref{helmholtz}) is 
\begin{equation}
\varphi=\frac{4\pi}{k^2}\rho_{0}\Bj_0(kr).
\end{equation}
This yields the electric fields
\begin{eqnarray}
\mathbf{E}_L  =& -\nabla\phi &=  \frac{4\pi}{k}\rho_{0}j_1(kr)\mathbf{n}, \\
\mathbf{E}_D  =& -\mathbf{E}_L &=  -\frac{4\pi}{k}\rho_{0}j_1(kr)\mathbf{n}.
\end{eqnarray}
The effective gravitational source is therefore
\begin{eqnarray}
t_{00} &=& -\frac{1}{4}\gamma_{[\rho\sigma],\kappa}\gamma^{[\rho\sigma],\kappa}
-\frac{1}{2}\gamma_{[\rho\sigma],\kappa}\gamma^{[\rho\kappa],\sigma}
+\frac{1}{2}\gamma^{[\rho\sigma]}\nabla^2\gamma_{[\rho\sigma]}\\
&=& G\left(\frac{4\pi \rho_{0}}{c^2k}\right)^2\left[-\frac{1}{6}(\Bj_0(kr))^2-(\Bj_1(kr))^2+\frac{2}{3}(\Bj_2(kr))^2\right].\nonumber
\end{eqnarray}

Solving $\nabla\gamma_{00}=-t_{00}$ with the condition that its derivative
vanishes at spatial infinity we obtain
\begin{equation}
\gamma_{00}=G\left(\frac{2\pi \rho_{0}}{c^2k^2}\right)^2[\ln(2kr)-\Ci(2kr)+\Bj_0(2kr)-2(\Bj_1(kr))^2].
\end{equation}
This results in an acceleration of 
\begin{equation}
\mathbf{a}=-G\left(\frac{\pi \rho_{0}}{ck^2}\right)^2[1-\Bj_0(2kr)-4\sin(kr)\Bj_1(kr)+8\Bj_1^2(kr)]\frac{\mathbf{n}}{r}.
\end{equation}
Note the log term in $\gamma_{00}$ cancels with the cosine integral at the
origin to yield a finite result, and the acceleration vanishes at the origin.  

As before this can be fit to a flat galactic rotation curve by requiring
\begin{equation}
G\left(\frac{\pi \rho_{0}}{ck^2}\right)^2=v^2.
\end{equation}
And as before we can check whether the acceleration is negligible near the
central black hole.  This will depend on $\ell$, but some analysis shows that
$a_r$ is bounded by $2G\left(\frac{\pi \rho_{0}}{ck^2}\right)^2/r=2v^2/r$
independently of $\ell$.  This ensures that $|a_r|$ is less than 10\%
the acceleration due to Sgr~A* for $r<3000$~AU in the Milky Way, for example.

A difficulty with this particular solution, however, is its nonzero (in fact maximum) 
charge density
at the origin. 
Unlike the axially symmetric solution investigated previously,  
this charge density does not vanish when integrated over a spherical volume.
A central black hole, then, must either be allowed to acquire charge, 
or the central charge must be removed.  In either case the net effect would be
the addition of a large, monopole electric field, which might destabilize 
the surrounding charge distribution.

\acknowledgments

The author gratefully acknowledges the late C.~R. Johnson 
for many valuable discussions, including
the conjecture that motivated this paper,
as well as for some of the mathematical content of
Section III.A and the Appendix.
The author also thanks S. Carlip and L. Magnani for helpful discussions.

\bibliographystyle{unsrt}
\bibliography{sphereless}

\end{document}